\documentclass[preprint,aps,floats,showpacs,amssymb,tightenlines]{revtex4}
\usepackage{epsfig}

\newcommand{\be}{\begin{equation}}
\newcommand{\ee}{\end{equation}}
\newcommand{\bea}{\begin{eqnarray}}
\newcommand{\eea}{\end{eqnarray}}
\newcommand{\bml}{\begin{mathletters}}
\newcommand{\eml}{\end{mathletters}}

\begin{document}

%\preprint{IUB-TH-}

%\twocolumn[\hsize\textwidth\columnwidth\hsize\csname @twocolumnfalse\endcsname

%%%%%%%%%%%%%%%%%%%%%%%%%%%%%%%%%%%%%%%%%%%%%%%%%%%%%%%%%%%%%%%%%%%%%%%%%%

%\wideabs{    % Uncomment this line for two-column output

\title{
Localising gravity in composite monopole brane worlds without bulk cosmological
constant}

\author{Eug\^enio R. Bezerra de Mello\footnote{emello@fisica.ufpb.br}}
\affiliation{Departamento de F\'{\i}sica-CCEN, Universidade Federal da 
Para\'{\i}ba, 58.059-970, J. Pessoa, PB, C. Postal 5.008, Brazil}
\author{Betti Hartmann\footnote{b.hartmann@iu-bremen.de}}
\affiliation{School of Engineering and Science, International University Bremen (IUB),
28725 Bremen, Germany}
\date{\today}
%\setlength{\footnotesep}{0.5\footnotesep}

%%%%%%%%%%%%%%%%%%%%%%%%%%%%%%%%%%%%%%%%%%%%%%%%%%%%%%%%%%%%%%%%%%%%%%%%%%
\begin{abstract}
We study  a 7-dimensional brane world scenario with a Ricci-flat 3-brane
residing in the core of a composite monopole defect, i.e., a defect composed
of a 'tHooft-Polyakov and a global monopole. 
Admitting a direct interaction between the two bosonic sectors of the 
theory, we analyse the structure of the space-time in the limits of small, 
respectively  large direct interaction coupling constant.
For large direct interaction, the global monopole 
disappears from the system and leaves behind a negative cosmological constant
in the bulk such that gravity-localising solutions are possible without
the a priori introduction of a bulk cosmological constant. 
\end{abstract}

\pacs{04.20.Jb, 04.40.Nr, 04.50.+h, 11.10.Kk }
\maketitle

%%%%%%%%%%%%%%%%%%%%%%%%
\section{Introduction}
%%%%%%%%%%%%%%%%%%%%%%%%
Brane worlds scenarios have attracted much interest over the past years 
\cite{akama,dvali,anton,arkani,rs1,rs2}. These assume our world to be a 
$3-$brane, i.e. a $(3+1)-$dimensional submanifold embedded in a higher 
dimensional space-time. While the standard matter fields are confined to the 
brane, gravity lives in all dimensions. 
The idea that matter is confined to a lower
dimensional manifold is not a new one. The localization of fermions on a 
domain wall has been discussed in \cite{ruba}. Recently, it was newly motivated 
by results from string theory. In type I string theory so-called Dp-branes
exist on which open strings, which represent matter fields, end. 
Gravitational fields, which are represented by closed strings, 
live in the full dimensions. However, as is well known, Newton's law in 
four dimensions is well tested down to $0.2$ mm. Thus appropriate brane 
world models should localise gravity ``well enough'' to the 3-brane. 
In the so-called Randall-Sundrum (RS) models 
\cite{rs1,rs2} the space-time contains two (RSI), respectively one (RSII)
Ricci-flat 3-brane(s) 
embedded in a 5-dimensional Anti-de-Sitter (AdS) bulk. For the
localisation of gravity in these models, the brane tension
has to be fine-tuned to the negative bulk cosmological constant.
The RSI model proposes a possible solution to the 
hierarchy problem, which can be solved  
if the distance between the two branes is about $37$ times the AdS radius.

Although topological defects have been mainly studied in a 
four-dimensional space-time \cite{vilenkin_shellard}, they have recently been considered in the context of brane worlds. The localisation of gravity on different topological defects has been discussed \cite{shapo}. This includes domain walls \cite{dfgk}, Nielsen-Olesen strings \cite{shapo1,shapo2}, monopoles (both global \cite{bc} and local \cite{shapo3}) in $5$, $6$ and $7$  space-time dimensions, respectively. The 3-brane sits at the core of these higher dimensional topological defects. It was found \cite{shapo,shapo1}
that gravity-localising (so-called ``warped'') solutions are possible if certain relations between the defect's tensions hold. While in the case of domain walls and strings, gravity can only be localised if the bulk cosmological constant is negative, for magnetic monopoles the gravity-localisation is possible for both signs of the cosmological constant.

The seven-dimensional monopoles with a 3-brane residing at their core have also been discussed in \cite{Cho1} for a global monopole and in \cite{Cho2} for a local monopole without bulk cosmological constant.
The emphasis was put on the structure of space-time depending on the value of the symmetry breaking scale. Static solutions were found to be singular for the symmetry breaking scale being larger than the higher dimensional
Planck's constant. In the case of global monopoles this singularity appears because the deficit angle of the transverse space becomes larger than $4\pi$. The four-dimensional analogues of these solutions
have been called ``supermassive'' monopoles \cite{liebling}. The singularities were shown to be removable by letting the defects' cores inflate.

Composite topological defects have been first studied in a four-dimensional 
space-time. Specifically a composite monopole, i.e., a system composed by 
a 'tHooft-Polyakov and a global monopole, was analyzed in 
\cite{Mello,bh,bbh} and  more recently in \cite{ahu} (for  
composite strings see \cite{bbh1}). In \cite{ahu} the large direct interaction
between the two sectors of the theory was studied and it has been found that
for large enough direct interaction, the global monopole will disappear from
the system and leave behind a negative cosmological constant. 
The problem of stability of composite monopoles has been analyzed in 
\cite{Ana} and \cite{Mello2}.

In this paper, we try to join both strategies by studying a composite monopole system of a 'tHooft-Polyakov and a global monopoles in a seven-dimensional space-time with a 
3-brane residing at the core of the composite
system. Our bulk cosmological constant is zero. This is the seven-dimensional analogue of the solutions 
studied previously \cite{Mello,bh,bbh,ahu}. For small direct interaction between the 
global and the gauge sector, 
both defects are present and we investigate the solutions in the spirit of 
\cite{Cho1,Cho2}. For large direct interaction, the global monopole disappears from the system 
and ``leaves behind'' a negative
cosmological constant. The system then corresponds to the one studied in \cite{shapo3}.

Our paper is organised as follows: in Section II, 
we give the model including the ansatz 
and the equations of motion. 
In Section III, we discuss solutions for small direct interaction, i.e. 3-branes in the cores of composite monopoles, while in section IV we concentrate on large direct interactions with gravity localising solutions. We give our conclusions in section V.

%%%%%%%%%%%%%%%%%%%
\section{The Model}
%%%%%%%%%%%%%%%%%%%
In the following, we shall consider a composite monopole system in a seven-dimensional space-time. 
The composite defect resides in the three dimensional transverse submanifold and has its core at a $(3+1)-$dimensional world-volume.

The seven-dimensional action associated with this model reads~:
\begin{equation}
S=\int d^7x\sqrt{-g}\left(\frac{R}{16\pi G_N}+{\cal{L}}_m\right) \ ,
\end{equation}
where $G_N$ is the seven-dimensional gravitational constant with $G_N=M_{pl(7)}^{-5}$ and $M_{pl(7)}^{-5}$ denotes the 
$7$-dimensional Planck constant.

The matter field Lagrangian density in terms of the 
Higgs field $\phi^a$, the gauge field $A_M^a$ and the
Goldstone field $\chi^a$ is given by~:
\begin{equation}
\label{L}
{\cal L}_m =-\frac14 F^a_{MN}F^{a,MN}-\frac12(D_M\phi^a)(D^M\phi^a)-
\frac12(\partial_M \chi^a)(\partial^M\chi^a)-
V(\phi^a,\chi^a) \ , \ a=1,2,3 \ 
\end{equation}
with the covariant derivative of the Higgs field $\phi^a$~:
\begin{equation}
D_M\phi^a=\partial_M\phi^a-e\epsilon_{abc}A_M^b\phi^c \ ,
\end{equation}
the field strength tensor
\begin{equation}
F_{MN}^a=\partial_M A_N^a-\partial_N A_M^a-e\epsilon_{abc}A_M^b A_N^c \ ,
\end{equation}
and the potential
\begin{eqnarray}
V(\phi^a,\chi^a)&=&\frac{\lambda_1}4\left(\phi^a\phi^a-\eta_1^2\right)^2+
\frac{\lambda_2}4\left(\chi^a\chi^a-\eta_2^2\right)^2\nonumber\\
&+&\frac{\lambda_3}2(\phi^a\phi^a-\eta_1^2)(\chi^a\chi^a-\eta_2^2) \ .
\end{eqnarray}
In the system, $e$ denotes the gauge coupling, $\lambda_1$ and $\lambda_2$ the self-couplings
of the Higgs, respectively Goldstone field, while $\eta_1$ and $\eta_2$ 
the respective vacuum expectation values of the Higgs and Goldstone fields. The parameter
$\lambda_3$ is the direct interaction 
coupling constant between the global and local sectors 
discussed previously in four-dimensional models \cite{bbh,ahu}.

%%%%%%%%%%%%%%%%%%%%%%%%%%
\subsection{The ansatz}
%%%%%%%%%%%%%%%%%%%%%%%%%%
The most general static seven-dimensional metric tensor with spherical symmetry in the 
extra three dimensions is given by \cite{Cho1,Cho2}:
\begin{equation}
\label{7d}
d\hat{s}^2=M^2(r){\bar{g}}_{\mu\nu}dx^\mu dx^\nu
+dr^2+C^2(r)r^2\left(d\theta^2+\sin^2\theta d\phi^2\right)=g_{AB}dx^Adx^B \ 
\end{equation}
with $A$, $B=0,... \ ,6$ and $\mu$, $\nu=0,... \ ,3$. In the line element above
${\bar{g}}_{\mu\nu}$ represents the metric of the $(3+1)$-dimensional world-volume. 
In the following, we shall assume that the 3-brane is Ricci flat, i.e.
$\bar{g}_{\mu\nu}=\eta_{\mu\nu}$. 
Here we
use the following notation for the coordinates: 
$x^A=(x^\mu,x^i)=(x^\mu,r,\theta,\phi)$.

For the Higgs and Goldstone fields, we use the standard hedgehog Ans\"atze~:
\begin{equation}
\label{Higgs1}
\chi^a(x)=\eta_1f(r) e_r^a \ \ , \ \
\phi^a(x)=\eta_1 h(r)e_r^a \ ,
\end{equation}
while for the gauge fields, we choose~:
\begin{equation}
A_{\theta}^a(r)=-\frac{1-u(r)}{e} e_{\phi}^a \ \  , \ \
A_{\phi}^a(r)=\frac{1-u(r)}{e} \sin\theta e_{\theta}^a \ \  ,
\end{equation}
and
\begin{equation}
\label{A0}
A_r^a(r)=0 \ \  , \ \  A_{\mu}^a(r)=0  \ .
\end{equation}
The Lagrangian density (\ref{L}) is terms of these Ans\"atze then reads~:
\begin{eqnarray}
\label{lagrange}
{\cal{L}}_m&=&-\frac12\eta_1^2(f')^2-\frac12\eta_1^2(h')^2
-\frac{(u')^2}{e^2C^2r^2}-\frac{(u^2-1)^2}{2e^2C^4r^4}
-\frac{\eta_1^2 u^2 h^2}{C^2r^2}-\frac{\eta_1^2f^2}{C^2r^2}\nonumber\\
&-&\frac{\lambda_1\eta_1^4}4(h^2-1)^2
-\frac{\lambda_2\eta_1^4}4(f^2-q^2)^2-\frac{\lambda_3\eta_1^4}2(h^2-1)(f^2-q^2) \ 
\end{eqnarray}

%%%%%%%%%%%%%%%%%%%%%%%%%%%%%%%%%
\subsection{Equations of Motion}
%%%%%%%%%%%%%%%%%%%%%%%%%%%%%%%%%%
We introduce the dimensionless variables $x=e\eta_1 r$
such that the equations of motion depend only on the following coupling constants:
\begin{equation}
q=\frac{\eta_2}{\eta_1} \ \ , \ \ \alpha^2=8\pi G_N \eta_1^2 \ \ , \ \ \beta_i=\frac{\lambda_i}{e^2} \ , \ i=1,2,3  \ .
\end{equation}
The seven dimensional Einstein equations
\begin{eqnarray}
G_A^B=8\pi G_NT_A^B \ ,
\end{eqnarray}
with the energy-momentum tensor
\begin{eqnarray}
T^B_A=F_{AC}^aF^{{a}BC}+(D_A\phi^a)(D^B\phi^a)+(\partial_A\chi^a)(\partial^B\chi^a)+\delta^B_A{\cal{L}}_m \ ,	
\end{eqnarray}
give the following set of non-linear coupled differential equations (the prime denotes here and in the following the 
derivative with respect to $x$):
\begin{equation}
\label{einstein1}
3\frac{M''}{M}+3\frac{M'^2}{M^2}+6\frac{M'}{Mx}+6\frac{M'C'}{MC}+2\frac{C''}{C}+6\frac{C'}{Cx}+\frac{C'^2}{C^2}+\frac1{x^2}-\frac1{C^2x^2}=\alpha^2 T^0_0 \ ,
\end{equation}
\begin{equation}
\label{einstein2}
8\frac{M'C'}{MC}+6\frac{M'^2}{M^2}+8\frac{M'}{Mx}+\frac{C'^2}{C^2}+2\frac{C'}{Cx}+\frac1{x^2}-\frac1{C^2x^2}=\alpha^2 T^x_x \ ,
\end{equation}
\begin{equation}
\label{einstein3}
4\frac{M''}{M}+ \frac{C''}{C}+2\frac{C'}{Cx}+4\frac{M'C'}{MC}+4\frac{M'}{Mx}+6\frac{M'^2}{M^2}=\alpha^2 T^{\theta}_{\theta} 
\end{equation}
with the non-vanishing components of the energy-momentum tensor
\begin{eqnarray}
T^{\mu}_{\mu} &=& 
-\frac{u'^2}{C^2x^2}-\frac{(1-u^2)^2}{2 C^4x^4}-\frac{1}{2} 
h'^2-\frac{1}{2} f'^2-\frac{h^2 u^2}{C^2x^2} 
-\frac{f^2}{C^2x^2}-{\cal{U}}(h,f) \ , \ \  \ \mu=0, 1, 2, 3 \ ,  \nonumber \\
T^x_x &=& \frac{u'^2}{C^2x^2}-\frac{(1-u^2)^2}{2C^4x^4}+\frac{1}{2}h'^2+\frac{1}{2} f'^2-\frac{h^2 u^2}{C^2x^2}-\frac{f^2}{C^2x^2}-{\cal{U}}(h,f)  \ ,   \nonumber \\
T^{\theta}_{\theta} &=& \frac{(1-u^2)^2}{2 C^4x^4}- \frac{1}{2} h'^2 - \frac{1}{2} f'^2-{\cal{U}}(h,f) \ ,
\end{eqnarray}
and the potential
\begin{eqnarray}
{\cal{U}}(h,f)=	\frac{\beta_1}{4}(h^2-1)^2+\frac{\beta_2}{4}(f^2-q^2)^2+\frac{\beta_3}{2}(h^2-1)(f^2-q^2) \ .
\end{eqnarray}

The conservation of the energy-momentum tensor, $D_MT^M_N=0$, implies
\begin{eqnarray}
\frac d{dx}T^x_x=4\frac{M'}M\left(T^0_0-T^x_x\right)+\left(\frac{C'}C+\frac1x\right)\left(T^\theta_\theta-T^x_x\right) \ ,
\end{eqnarray} 
consequently equations (\ref{einstein1}) to (\ref{einstein3}) are functionally dependent \cite{shapo}.

Combining conveniently the above equations, two differential equations for the unknown metric functions $M$ and $C$ are obtained. 
The first combination gives:
\begin{equation}
\label{A1}
\frac{(C^2x^2 M' M^3)'}{C^2x^2 M^4}=\frac{\alpha^2}{5}\left(\frac{(1-u^2)^2}{C^4x^4}+2\frac{u'^2}{C^2x^2}- 2{\cal{U}}(h,f)\right) \ ,
\end{equation}
while the second reads:
\begin{eqnarray}
\label{A2}
\frac{[M^4 (C^2x+C'Cx^2)]'}{C^2x^2 M^4}-\frac1{C^2x^2}&=&\frac{\alpha^2}{5}\left(\frac{-3u'^2}{C^2x^2}-
\frac{4(1-u^2)^2}{C^4x^4}-\frac{5 h^2 u^2}{C^2x^2}\right.\nonumber\\
&-&\left.5\frac{f^2}{C^2x^2}-2{\cal{U}}(h,f)\right)  \ .
\end{eqnarray}

Varying $S_m$ with respect to the matter fields we obtain the Euler-Lagrange equations for the Higgs field
function:
\begin{equation}
\label{higgseq}
\left(M^4 C^2x^2 h'\right)'=M^4 C^2x^2\left(2\frac{u^2 h}{C^2x^2} 
+ \beta_1 h (h^2-1) + \beta_3 h (f^2-q^2)\right) \ ,
\end{equation}
for the gauge field function
\begin{equation}
\label{gaugeeq}
\left(M^4 u'\right)'=M^4\left(-\frac{u(1-u^2)}{C^2x^2} + h^2 u\right) \ ,
\end{equation}
and for the Goldstone field function
\begin{equation}
\left(M^4C^2x^2 f' \right)'=M^4f\left(2+\beta_2C^2x^2(f^2-q^2) + \beta_3C^2x^2(h^2-1) \right)  \ .
\end{equation}

It has been noticed in \cite{bbh,ahu} that the behavior of the composite system in four dimensional space-time depends strongly on the combination of the coupling constants appearing in the potential
\begin{equation}
\Delta:=\beta_1 \beta_2-\beta_3^2  \ .
\end{equation}
The case $\Delta > 0$ was studied in detail in \cite{bbh}. For $\Delta > 0$,
the minimum of the potential is given by
\begin{equation}
h(x)=1  \ , \ f(x)=q
\end{equation} 
such that asymptotically $h(x\rightarrow \infty)\rightarrow 1$ and
$f(x\rightarrow \infty)\rightarrow q$. 

The case $\Delta \leq 0$ was reinvestigated in \cite{ahu} and it was found
that the boundary conditions at infinity have to be generalized for the
Higgs and Goldstone field functions. The reason for this is that there are two possible minima for the potential:
\begin{equation}
\label{exotic}
{\rm (a)}  \ \ h^2(x)=0 \ , \ f^2(x)=\left(q^2+\frac{\beta_3}{\beta_2}\right) 
\ \ \ {\rm or}  \ \ {\rm (b)} \ \
f^2(x)=0 \ , \ h^2(x)=\left(1+\frac{\beta_3}{\beta_1}q^2\right) \ .   
\end{equation}
We give the corresponding boundary conditions for the two cases
in Sections III and IV, respectively.

%%%%%%%%%%%%%%%%%%%%%%%%%%%%%%%%%%%%%%%%%%%%%%%%%%%%%%%%%%%%%%%%%%%%%%%%%%%%%%%%
\section{$\Delta > 0$ solutions: 3-branes in the core of composite monopoles}
Here we shall analyse the system in 
the regime of small direct interaction coupling. 
For this case, both the local and the global monopole are present.
%%%%%%%%%%%%%%%%%%%%%%%%%%%%%%%%%%%%%%%%%%%%%%%%%%%%%%%%%%%%%%%%%%%%%%%%%%%%%%%%%
\subsection{Analytic solutions}
%%%%%%%%%%%%%%%%%%%%%%%%%%%%%%%%%%%%
Before to embark in a numerical analysis, let us first investigate possible
analytical solutions for the system of equation.
Setting the matter functions equal to their vacuum values, i.e.,
$h(x)\equiv 1$, $f(x)\equiv q$ and $u(x)\equiv 0$ in the Lagrangian 
(\ref{lagrange}), we can find analytic solutions
of the system determined by the equations (\ref{A1}) and (\ref{A2}) alone. 
These are by themselves of interest, but we would also
expect our solutions to tend to these solutions very far away ($x\to\infty$) 
from the core of the composite monopole. Analytic solutions for brane world space-times
have been investigated for the ``pure'' global monopole case for points outside the monopole's core in \cite{ov}.

Here, we introduce two new functions $w(x):=M^4(x)$ and 
$v(x)=:x^2 C^2(x)$. The equations (\ref{A1}) and (\ref{A2}) can then be written as follows:
\begin{eqnarray}
(w'v)'&=&\frac{4\alpha^2}5\frac wv \\
(wv')'&=&2(1-q^2\alpha^2)w-\frac{8\alpha^2}5\frac wv \ .
\end{eqnarray}
It is possible to obtain the differential equation obeyed by $v$ only. It is:
\begin{eqnarray}
5v''v\mp 2\alpha\sqrt{5}v'-10(1-\alpha^2q^2)v+8\alpha^2=0 \ .
\end{eqnarray}
This differential equation is a non-linear one, consequently no general solutions could be
found by us. In the following, we shall try to find particular analytic solutions to these equations.
As a first trial, using the Ansatz $v(x)=constant$, we found the 
following solution:
\begin{eqnarray}
v(x)\equiv \frac{4\alpha^2}{5(1-\alpha^2q^2)} 
\end{eqnarray}
and 
\begin{eqnarray}
w(x)=C_1e^{\kappa x}+C_2e^{-\kappa x}	
\end{eqnarray}
with 
\begin{eqnarray}
\kappa=\frac{\sqrt{5}(1-\alpha^2q^2)}{2\alpha} \ ,
\end{eqnarray}
$C_1$, $C_2$ being two arbitrary constants.
This corresponds to a cylinder-like solution with the constant radius
of the 2-spheres equal to $2\alpha/\sqrt{5\left(1-\alpha^2 q^2\right)}$,
and holds also for the ``pure'' magnetic monopole case, $q=0$, which
has been studied in \cite{Cho2}.

In the case of ``pure'' global monopole, it has been found in \cite{ov} that the constant radius of the 2-spheres is 
arbitrary, and the solution $M^4=constant$ exists only for $\alpha^2 q^2=1$. We find in contrast 
to that for the composite case, that the radius of the 2-sphere
is fixed by the coupling constants, and the solutions exist
for all combinations of $\alpha$ and $q$ as long as $\alpha^2 q^2 < 1$ and that $M^4$ behaves exponentially.

Note that for $C_1=0$, this solution would have the form of a gravity-localising one \cite{shapo}. However, 
in practice, i.e., when integrating the equations
numerically, we shall not find these type of solutions in the asymptotic region
of the space-time since the assumption $f(x)\equiv q$ holds only very far 
from the core of the composite ($f$ falls of power-like), however we 
always integrate on finite intervals.

The solution  $w(x)=constant$ that has been found for the ``pure'' global
monopole case \cite{ov}, is not a solution to the equations (\ref{A1}) and (\ref{A2}).
However, for points very far from the monopole's core, i.e., for $x \gg 1$,  we can neglect the $1/(C x)^4$ term in (\ref{A1}) and (\ref{A2}) 
and we obtain the set of equations:
\begin{eqnarray}
(vw')'&=&0 \ , \\
(wv')'&=&2(1-\alpha^2q^2)w \ .	
\end{eqnarray}
This system has the following possible solution:
\begin{eqnarray}
v(x)&=&(1-\alpha^2q^2)x^2  \\
w(x)&=&C_3 \ ,	
\end{eqnarray}
$C_3$ being constant. The line element associated with this solution (choosing $C_3=1$) 
is then given by:
\begin{eqnarray}
ds^2=\eta_{\mu\nu}dx^\mu dx^\nu+dr^2+(1-\alpha^2q^2)r^2\left(d\theta^2+\sin^2\theta d\phi^2\right) \ .	
\end{eqnarray}
This is nothing else but a space-time with solid angle deficit in the three dimensional transverse submanifold,
equal to the solution found for the ``pure'' global monopole case in \cite{ov}.  
This result is not surprising since we would expect the global monopole
to dominate the system far from the core of the composite.

%%%%%%%%%%%%%%%%%%%%%%%%%%%%%%%%
\subsection{Boundary conditions}
%%%%%%%%%%%%%%%%%%%%%%%%%%%%%%%%%
For the numerical construction of the solutions, we have to impose boundary conditions.
We require regularity of the solutions at the origin and thus impose:
\begin{equation}
\label{bc1}
M(x=0)=1 \ , \ M'(x)|_{x=0}=0 \ , \ C(x=0)=1 \ , 
\end{equation}
\begin{equation}
\label{bc2}
h(x=0)=0 \ , \ u(x=0)=1 \ , \ f(x=0)=0 \ ,
\end{equation}
while the requirement of finite energy leads to
\begin{equation}
\label{bc3}
h(x)|_{x\rightarrow \infty}=1 \ , \ u(x=\infty)=0 \ , \ f(x)|_{x\rightarrow \infty}=q   \ .
\end{equation}

%%%%%%%%%%%%%%%%%%%%%%%%%%%%%%
\subsection{Numerical results}
%%%%%%%%%%%%%%%%%%%%%%%%%%%%%%%

We have studied the case $\Delta > 0$ first, and followed in our investigation the studies in \cite{Cho1, Cho2}.
First, we have fixed $\beta_i$, $i=1,2,3$ as well as $\alpha$ to some value and studied
the development of the solutions for increasing $q$. 

\begin{figure}[!htb]
\centering
\leavevmode\epsfxsize=9.0cm
\epsfbox{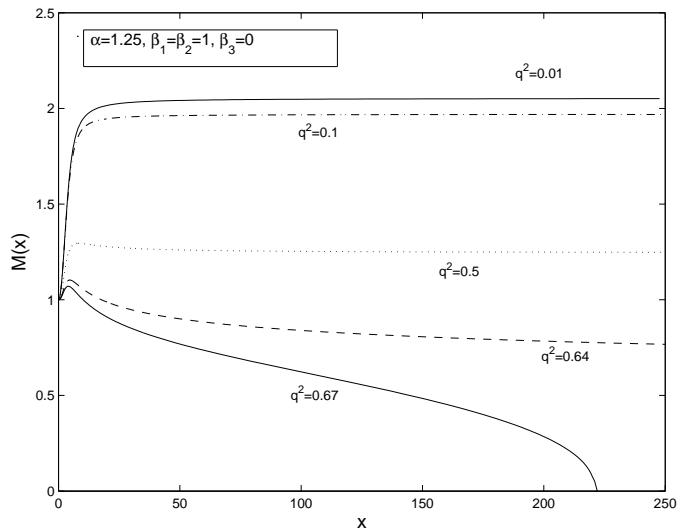}\\
\caption{\label{fig1} The profile of the 
metric function $M(x)$ for the composite monopole system without direct
interaction ($\beta_3=0$) and $\beta_1=\beta_2=1$, $\alpha=1.25$ and
different values of the ratio of vacuum expectation values $q=\eta_2/\eta_1$.}
\end{figure}

\begin{figure}[!htb]
\centering
\leavevmode\epsfxsize=9.0cm
\epsfbox{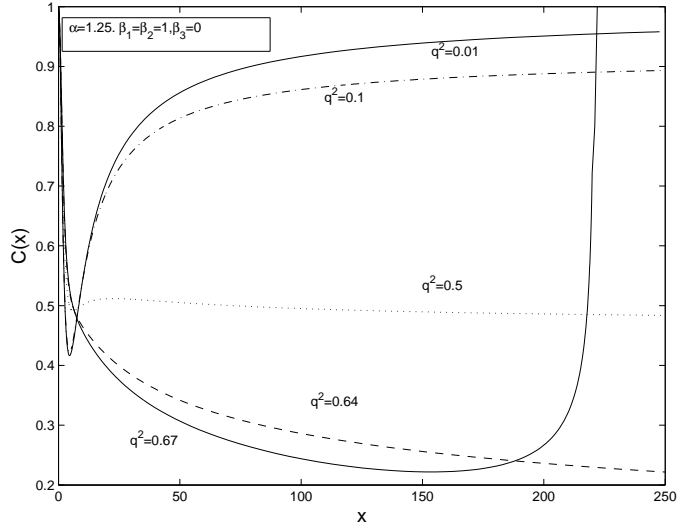}\\
\caption{\label{fig2} 
Same as Fig.\ref{fig1} for the metric function $C(x)$.}
\end{figure}

Our results for $\beta_1=\beta_2=1$, $\beta_3=0$ and $\alpha=1.25$ are shown in Fig.s \ref{fig1} and \ref{fig2}.
Asymptotically the behavior of the metric functions is very similar to that observed in the case
of global monopoles \cite{Cho1}, which is not surprising since the
global monopole -in contrast to the local one - has long range fields. The function
$C(x)$ tends to a value smaller than unity, indicating that the space-time has
a solid deficit angle which increases with increasing $q$ and thus increasing $\eta_2$.
For $\alpha^2 q^2 =1$ the deficit angle becomes equal to $4\pi$. Increasing $q^2$ beyond its critical value,
$q_c^2=1/\alpha^2$, leads to a curvature singularity of the solutions with the metric function
$M(x)$ vanishing at $x=x_s(q_c)$ and $C(x)$ tending to infinity at this point \footnote{It has been shown that 
the functions $M$ and $C$ present opposite singular behavior for pure
global \cite{Cho1} and local \cite{Cho2} monopole systems, respectively. 
Here, for the composite monopole system, we show that
its singular behavior is similar to that of the global monopole one}.
We show the solution for $q^2=0.67$ which has
a curvature singularity at $x\approx 220$. 
However, what is different here in comparison to the pure global case is that the metric
function $C(x)$ develops a minimum at small $x$. This is a typical feature of the
local monopole in curved space-time \cite{Cho2} (compare also the four-dimensional case \cite{bfm})
and is only present for small $q$. 

\begin{figure}[!htb]
\centering
\leavevmode\epsfxsize=9.0cm
\epsfbox{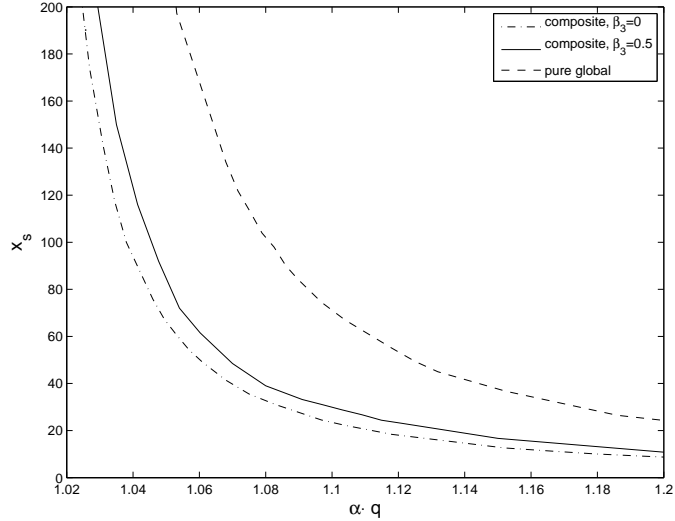}\\
\caption{\label{fig3} 
The location of the singularity at $x=x_s$ in dependence on the product
$\alpha\cdot q$ is shown for the composite monopole system
without direct interaction (dotted-dashed), the composite monopole
system with $\beta_3=0.5$ (solid) and for the pure global monopole
system (dashed).
}
\end{figure}

In Fig.\ref{fig3} we show the localisation of the singularity as function of $\alpha q$.  For comparison, we also show the corresponding values for the composite system
with direct interaction ($\beta_3=0.5$) and for the pure global monopole case.
For a fixed value of $\alpha q$, the value of $x_s$ is larger for the pure global
monopole in comparison to the composite system. This is related to the localisation
of energy close to the core (coming from the local monopole). Moreover, the increase of
$\beta_3$ increases the value of $x_s$ for a fixed $\alpha q$. 

%%%%%%%%%%%%%%%%%%%%%%%%%%%%%%%%%%%%%%%%%%%%%%%%%%%%%%%%%%%%%%%%%%%%%%%%%%%%%%%%%%%%%%%%%%%%%
\section{$\Delta < 0$ solutions: gravity localisation without bulk cosmological constant}
Now we consider the system for the regime of larger direct interaction coupling, i.e., $\lambda_3^2>\lambda_1\lambda_2$.
%%%%%%%%%%%%%%%%%%%%%%%%%%%%%%%%%%%%%%%%%%%%%%%%%%%%%%%%%%%%%%%%%%%%%%%%%%%%%%%%%%%%%%%%%%%%%%
\subsection{Exotic composites}
%%%%%%%%%%%%%%%%%%%%%%%%%%%%%%%%%
For $\Delta < 0$, one of the defects disappears from the system, 
like in the four-dimensional case, see \cite{ahu}, 
while the scalar function of the remaining defect tends 
asymptotically to the respective value given in (\ref{exotic}).
The space-time then becomes asymptotically Anti-de-Sitter with an
effective {\it negative} cosmological constant which reads:
\begin{equation}
{\rm For \ (a)} \ \ \Lambda_{(a)}:= \frac{\eta_1^4e^2}{4\beta_2} \Delta   \ \ .  
\ \ \ {\rm For \ (b)}  \ \ \Lambda_{(b)}:=\frac{q^4\eta_1^4e^2}{4\beta_1} \Delta 
\end{equation}
In our system, for $\Delta < 0$ and $f(x)\equiv 0$, the global monopole disappears since this minimises the energy, consequently it 
corresponds to the system analysed in \cite{shapo3}, in which gravity-localising solutions of a 'tHooft-Polyakov
monopole in seven dimensions have been studied. In the following, we shall
discuss the system studied here in comparison to that given in \cite{shapo3}.

The metric functions of the gravity-localising
solutions read \cite{shapo3}:
\begin{equation}
M=M_0 e^{-c_1 x/2} \ \ , \ \ C(x)=c_2/x 
\end{equation} 
with the constants $c_1$ and $c_2$ that can be computed directly from
(\ref{A1}) and (\ref{A2}) (with $\gamma_{(b)}=\frac{\Lambda_{(b)}}{e^2\eta_1^4}$) :
\begin{eqnarray}
\label{coeff}
c_1^2&=&\frac{5}{32 \alpha^2}\left(1-\frac{16}{5} \gamma_{(b)} \alpha^4 \pm \sqrt{1-\frac{32}{25} 
\gamma_{(b)} \alpha^4}\right)=\frac{5}{32 \alpha^2}\left(1-\frac{4}{5}\frac{q^4}{\beta_1} \Delta  \alpha^4 \pm 
\sqrt{1-\frac{8}{25}\frac{q^4}{\beta_1} \Delta   \alpha^4}\right) \, \nonumber \\
c_2^2&=&\frac{8\alpha^2}{5}\left(1\pm \sqrt{1-\frac{32}{25} \gamma_{(b)} \alpha^4}\right)^{-1}=\frac{8\alpha^2}{5}\left(1\pm \sqrt{1-\frac{8}{25} \frac{q^4}{\beta_1}\Delta \alpha^4}\right)^{-1} \ ,
\end{eqnarray}
Gravity localising solutions exist only for positive roots in $c_1^2 > 0$ and $c_2^2 > 0$, i.e., we use only the solutions of (\ref{coeff}) with the plus sign.

For the gauge and Higgs field functions, $u(x)$ and $h(x)$, we linearize the equations
(\ref{higgseq}) and (\ref{gaugeeq}) for $x \gg 1$ with
\begin{equation}
u(x)=\delta u(x) \ \ \ , \ \ \ h(x)=\sqrt{1+\frac{\beta_3}{\beta_1}q^2}-\delta h(x) \ .
\end{equation} 
Neglecting terms of order $\delta^2u$ and $\delta^2h$, we obtain:
\begin{eqnarray}
\delta u'' -2 c_1 \delta u' + \delta u \left( \frac{1}{c_2^2} - 1 
- \frac{\beta_3}{\beta_1} q^2\right)&=&0 \nonumber \\
\delta h'' - 2c_1 \delta h' - 2 \beta_1 \delta h \left( 1+ \frac{\beta_3}{\beta_1} q^2\right)&=&0 
\end{eqnarray}
Then setting $\delta u=u_0 e^{-k x}$ and $\delta h=h_0 e^{-mx}$, we find
\begin{eqnarray}
\label{parameters}
k_{\pm}&=&-c_1 \pm \sqrt{ c_1^2-c_2^{-2} + 1 + \frac{\beta_3}{\beta_1} q^2} \ \nonumber \\
m_{\pm}&=&-c_1 \pm \sqrt{c_1^2 + 2\beta_1 \left(1+\frac{\beta_3}{\beta_1} q^2\right) } \ .
\end{eqnarray}
These results agree with the ones in \cite{shapo3} for 
$\beta_3=0$, i.e. no direct interaction between the local and global sector
and/or $q=0$, i.e. a pure local monopole.

For $u$ and $h$ tending to finite values asymptotically, we have to require
$k > 0$ and $m > 0$. Since for gravity-localising solutions $c_1 > 0$, we can
only choose the solutions with the plus signs in (\ref{parameters}).
Then $c_2^{-2} < 1+ q \beta_3/\beta_1$ and with (\ref{coeff})
a relation between the couplings of the theory can be found. However, this is
not one unique solution, but a domain of solutions (see also below the ``Results''
section).

%%%%%%%%%%%%%%%%%%%%%%%%%%%%%%%%
\subsection{Boundary conditions}
%%%%%%%%%%%%%%%%%%%%%%%%%%%%%%%%%
For the numerical construction of the solutions, we have to impose boundary conditions.
The boundary conditions for the functions at the origin are equivalent to the ones given in
(\ref{bc1}) and (\ref{bc2}). However, the boundary conditions at infinity (\ref{bc3}) 
are now replaced by:
\begin{equation}
\label{bcinfty}
h'(x)|_{x\rightarrow \infty}=0 \ , \ u(x=\infty)=0 \ , \ f'(x)|_{x\rightarrow \infty}=0   \ .
\end{equation}
Note that the boundary conditions (\ref{bcinfty}) are the ones chosen in \cite{ahu} such that depending on the choice of $\beta_1$, $\beta_2$ and
$\beta_3$ the Higgs and Goldstone fields tend to their respective
minimum values.

%%%%%%%%%%%%%%%%%%%%%%%%%%%%%%%
\subsection{Results}
\begin{figure}[!htb]
\centering
\leavevmode\epsfxsize=9.0cm
\epsfbox{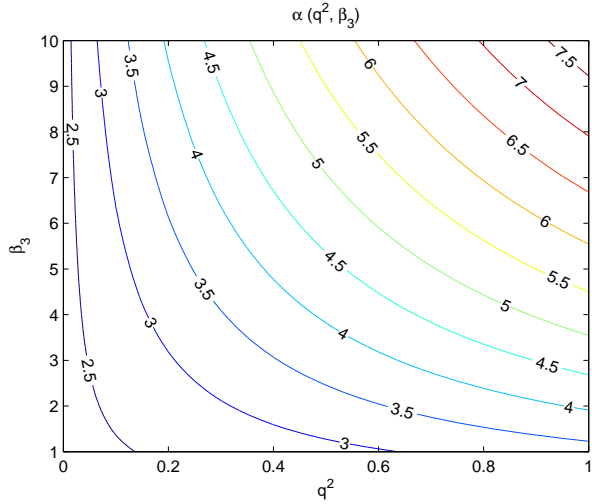}\\
\caption{\label{fig4} 
The contour plots of $\alpha(q^2,\beta_3)$ are given for gravity localising solutions with
$\beta_1=1.0$, $\tilde{\alpha}^2=5.5$ and $\tilde{\Lambda}\approx -0.054$.
}
\end{figure}

\begin{figure}[!htb]
\centering
\leavevmode\epsfxsize=9.0cm
\epsfbox{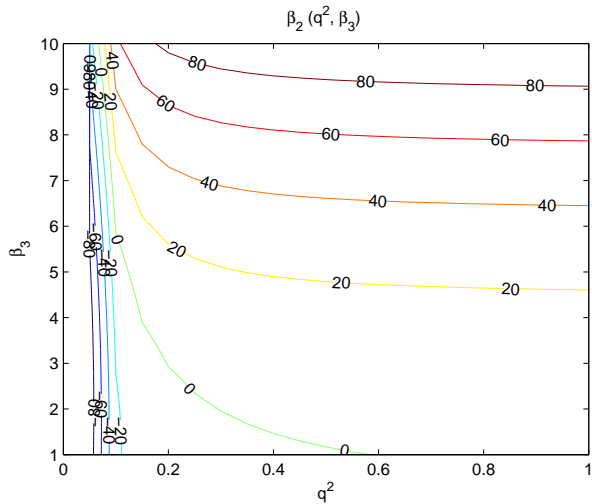}\\
\caption{\label{fig5} 
Same as Fig.\ref{fig4}, but for $\beta_2(q^2,\beta_3)$.
}
\end{figure}

%%%%%%%%%%%%%%%%%%%%%%%%%%%%%%
We have not performed explicit numerical calculations here, since 
our system of equations for $f\equiv 0$ 
and the non-vanishing functions $M$, $C$ and $\tilde{h}$ (with coupling constants
$\tilde{\alpha}$, $\tilde{\Lambda}$, $\beta_1$) is exactly 
equivalent to the one in \cite{shapo3}, if we define
\begin{equation}
\tilde{h}=\frac{h}{\sqrt{1+\frac{\beta_3}{\beta_1}q^2}} \ \ , \ \ 
\tilde{x}=x \sqrt{1+\frac{\beta_3}{\beta_1}q^2} \ \ , \ \ 
\tilde{\alpha}=\alpha \sqrt{1+\frac{\beta_3}{\beta_1}q^2} \ \ , \ \
\tilde{\Lambda}=\frac{q^4 (\beta_1\beta_2 - \beta_3^2)}{4 (\beta_1+\beta_3 q^2)}
\end{equation} 
with $\tilde{\Lambda} < 0$. Detailed numerical calculations have been
given there and we do not want to repeat them here. However, we would
like to point out the relation of the coupling constants of our theory to the
coupling constants given previously. In \cite{shapo1} it was found that gravity-localising solutions
exist only in a specific parameter domain in the $\tilde{\alpha}-\beta_1-\tilde{\Lambda}$-domain.
One solution found with negative bulk cosmological constant has values
$\beta_1=1.0$, $\tilde{\alpha}^2=5.5$ and $\tilde{\Lambda}\approx -0.054$. In our case,
fixing $\beta_1$, $\tilde{\alpha}^2$ and $\tilde{\Lambda}$, $\alpha$ and $\beta_2$ become 
functions of $\beta_3$ and $q^2$. The contour plots are given in Fig.\ref{fig4} and Fig.\ref{fig5}.
 
%%%%%%%%%%%%%%%%%%%%%%%
\section{Conclusions}
%%%%%%%%%%%%%%%%%%%%%%%
In this paper, we have studied a 7-dimensional brane world model with 
a composite monopole residing 
in the three dimensional transverse submanifold. 
We have considered a Ricci-flat 3-brane to be at the defect's core. 
Our analysis reveals that the 
geometrical structure of the three dimensional submanifold 
depends crucially on the parameter associated with the 
direct interaction between the local and the global sector of the theory. 
For small direct interaction, we found that 
qualitatively the system behaves asymptotically like 
the global monopole system representing a space-time with 
solid angle deficit in the transverse submanifold. 
The presence of the local magnetic monopole 
modifies the behavior of the metric functions only close to the defect's 
core.

For large  direct interaction, 
the global monopole naturally disappears from the 
system and leaves behind an effective negative cosmological constant. 
The system thus corresponds to a local monopole 
brane world scenario with negative bulk cosmological constant. 
Then, gravity localising solutions are possible without the a 
priori introduction of a bulk cosmological constant. 
In this context we were able to show that by a convenient 
redefinition of the matter fields and dimensionless coordinate 
that the present system is equivalent to the one previously 
analysed in \cite{shapo3}, 
consequently no explicit numerical calculations are needed.

%%%%%%%%%%%%%%%%%%%%%%%%%%%%%%%%%%%%%%%%%%%%%%%%%%%%%%%%%%%%%%%%%%%%

\end{document}